\documentclass[12pt]{article}

\usepackage{a41}
\usepackage{floatfig,epsfig}
\usepackage{rotating}

\newcommand{\la}[1]{\label{#1}}
\newcommand{\re}[1]{\ (\ref{#1})}
\newcommand{\nn}{\nonumber}
\newcommand{\ed}{\end{document}}
\newcommand{\be}{\begin{equation}}
\newcommand{\ee}{\end{equation}}

\newcommand{\ba}{\begin{eqnarray}}
\newcommand{\ea}{\end{eqnarray}}
\newcommand{\baz}{\begin{eqnarray*}}
\newcommand{\eaz}{\end{eqnarray*}}
\newcommand{\bb}{\begin{thebibliography}}
\newcommand{\eb}{\end{thebibliography}}
\newcommand{\ct}[1]{${\cite{#1}}$}

\begin{document}

\initfloatingfigs
\sloppy
\thispagestyle{empty}

\vspace{1cm}

\mbox{}

\vspace{5cm}

\begin{center}

{\large\bf   Pomeron  Fusion  and Central $\eta$ and $\eta^\prime$
Meson Production}\\
 N.I. Kochelev$^{1}$ , T.Morii$^2$, A.V. Vinnikov$^{3}$\\
\vspace{5mm}
{\small\it
(1) Bogoliubov Laboratory of Theoretical Physics,\\
Joint Institute for Nuclear Research,\\
 Dubna, Moscow region, 141980 Russia
\footnote{On leave of absence from Institute of Physics and Technology,
Almaty, 480082, Kazakhstan}
\\
(2) Faculty of Human Development, Division of Sciences for
Natural Environment\\
and Graduate School of Science and Technology,\\
Kobe University,
Nada, Kobe 657-8501, Japan \\
(3) Far Estern State University, Department of Physics,\\
Sukhanova 8, GSP, Vladivostok, 690660  Russia}
\vspace*{2cm}

\end{center}
\begin{abstract}
The contribution of pomeron fusion to the cross section
of $\eta$ and $\eta^\prime$ productions in double-diffractive
scattering has been calculated within the
Donnachie-Landshoff model of pomeron. It is shown that the double
pomeron exchange mechanism does not explain the full set of the
recent data of WA102 Collaboration, though it might not be
inconsistent with $\eta^\prime$ productions.
\end{abstract}
\newpage

 The structure of the pomeron is  now  widely under
discussion \ct{pom}.
One of the interesting ways to investigate this structure
is the double diffractive process(DDP).
Recently,  the precise experimental data on
central productions of pseudoscalar mesons,
$\pi^0$, $\eta$ and $\eta^\prime$, have been published
by WA102 Collaboration \ct{WA102}.
The most  spread wisdom is that the main contribution to the DDP
cross section should be related with the double pomeron exchange (DPE)\ct{DPE}.
The kinematics of DDP corresponds to a small momentum transfer region
and in this region the contribution from other processes such as
photon-photon and vector meson-vector meson fusion
is also possible \ct{gaga}.
However, the experimental data  give the value of the cross section
of pseudoscalar meson productions which are several orders of magnitude
larger than one can expect, for example, from $\gamma\gamma$ fusion \ct{WA102}.
One of the interesting feature of the data is unusual azimuthal
angle dependence of the cross section. Different mechanisms,  which
may responsible for an enhancement of the production cross section in the
kinematical region where the azimuthal angle between the $p_T$ vectors of
two final protons is 90 degrees,
have been discussed in recent papers \ct{gaga},\ct{close},
\ct{koch}.

The  subject of our paper is to estimate the DPE
contribution to the cross section of central $\eta$,
$\eta^\prime$ productions in WA102 kinematics.

A DPE diagram contributing to central $\eta$,
$\eta^\prime$ productions, is presented in Fig.1.
\begin{figure}[htb]
%\begin{figure}
\centering
\epsfig{file=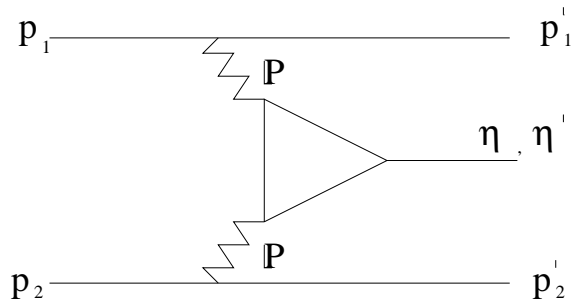,width=7.5cm}
\vskip 0.5cm
\caption{\it The DPE contribution to the  central pseudoscalar
meson  production}
\end{figure}
In order to estimate this contribution, the Donnachie-Landshoff(DL)
model of pomeron \ct{DL} will be used. In this model
the quark-quark interaction through the pomeron exchange is similar
to the photon exchange with modified propagator
\ba
M_{int}=i\beta_0^2\bar q(p_1^\prime)\gamma_\mu q(p_1)\bar q(p_2^\prime)
\gamma_\mu q(p_2) (S/S_0)^{\alpha_P(t)-1},
\la{int}
\ea
 where $S=(p_1+p_2)^2$, $t=(p_1-p_1^\prime)^2$, $\beta_0\approx 1.8 GeV^{-1}$,
$S_0\approx 1 GeV$ and $\alpha_P(t)=1+\epsilon+\alpha^\prime t$ is
the pomeron trajectory with $\epsilon\approx 0.085$ and $\alpha^\prime
=0.25 GeV^{-2}$.
To calculate the contribution of the diagram in Fig.1, one should
take into account  the form factors in pomeron-proton
and pomeron-pomeron-pseudoscalar meson vertices.
For the pomeron-proton vertex,   the electromagnetic form
factor of the nucleon
\be
 F_p(t)=\frac{4m_p^2-2.79t}{(4m_p^2-t)(1-t/0.71)},
\la{ff1}
\ee
where $m_p$ is the proton mass, is usually used \ct{DL}.
This form factor gives a rather good description of the $t$ dependence of the
elastic $pp$ cross section at high energy \ct{DL}.
Unfortunately, the pomeron-pomeron-pseudoscalar meson form factor
is not known well. Within the DL model, the property of the pomeron-quark
vertex  is similar
to the quark-photon vertex \ct{DL} and thus, for this form factor
we expect the same momentum transfer dependence as for the transition
pseudoscalar meson form factor $\gamma^*\gamma^*\rightarrow \eta,
\eta^\prime$. For a small momentum transfer which we consider here in DPE,
the Brodsky-Lepage  formula \ct{BL}
\be
F_{PPM}(t_1,t_2)=\frac{1}{(1-t_1/8\pi^2f_{PS}^2)(1-t_2/8\pi^2f_{PS}^2)}
\la{PPM}
\ee
can be used for this form factor, where $f_{PS}$ is a
meson decay constant which is related to partial width $\Gamma_{\gamma\gamma}$ (see \ct{kn})
\be
f_{PS}=\frac{\alpha}{\pi}\sqrt{\frac{M^3}{64\pi\Gamma_{\gamma\gamma}}}.
\label{fff}
\ee
The cross section of the meson production in the reaction
\be
p(p_1)+p(p_2)\rightarrow p(p_1^\prime)+p(p_2^\prime)+M(p_M)
\nn
\ee
is given by the formula
\be
d\sigma=\frac{dPS^3}{4\sqrt{(p_1.p_2)^2-m_p^4}}
\sum_{spin}|T|^2,
\la{cross}
\ee
where
\be
dPS^3=\frac{d^4p_1^\prime d^4p_2^\prime d^4p_M}{(2\pi)^9}(2\pi)^4
\delta(p_1^\prime-m_p^2)\delta(p_2^\prime-m_p^2)\delta(p_M^2-M^2)
\delta(p_1+p_2-p_1^\prime-p_2^\prime-p_M),
\la{ps}
\ee
is the 3-body phase space volume: $M$ is the meson mass,
$T$ is the matrix element for the DPE reaction and
$\sum_{spin}$ stands for spin summation and
spin average for the final and initial proton states, respectively.

At high energies and small momentum transfers, the four-momenta
of initial and final protons in the center of mass system are
given as
\begin{eqnarray}
p_1&\approx&(P+m_p^2/2P,\vec 0,P), {\ }{\ }{\ } {\ }
p_2\approx(P+m_p^2/2P,\vec 0,-P)\nn\\
{p_1}^\prime&\approx&(x_1P+(m_p^2+{\vec{p}_{1T}}^2)/2x_1P
,\vec{p}_{1T},x_1P),
{\ }{\ }{p_2}^\prime\approx(x_2P+{\vec{p}_{2T}}^2)/2x_2P,\vec{p}_{2T},-x_2P),
\label{kinematic}
\end{eqnarray}
where $P=\sqrt{S}/2$, $S=(p_1+p_2)^2$.
By using the result of Ref.\ct{frix} for the high energy phase
space volume at small momentum transfers, we obtain
\be
d^4p_{1,2}^\prime\delta(p_{1,2}^2-m_p^2)\approx \frac{1}{4}
dt_{1,2}dx_{1,2}d\Phi_{1,2},
\nn
\ee
where $\Phi_i$ are azimuthal angles of final protons and
$t_{1,2}=-q_{1,2}^2=(p_{1,2}-p_{1,2}^\prime)^2$.
Then, one can obtain
\be
dPS^3=\frac{1}{2^8\pi^4}dt_1dt_2dx_1dx_2d\Phi\delta(S(1-x_1)(1-x_2)-M^2)
\la{ps2}
\ee
for the phase space volume in the DPE reaction.
The matrix element is given by
\be
T=-9\beta_0^2F_p(t_1)F_p(t_2)\bar u(p_1^\prime)\gamma_\mu u(p_1)
\bar u(p_2^\prime)\gamma_\mu u(p_2)(S_1/S_0)^{\alpha_P(t_1)-1}
(S_2/S_0)^{\alpha_P(t_2)-1} p_M^\tau T_{\mu\nu\tau},
\la{matrix}
\ee
where $S_{1,2}=(q_{2,1}+p_{1,2})^2$, $S_0\approx 1 GeV$ and
$T_{\mu\nu\tau}$
is the matrix element for two pomeron fusion
into pseudoscalar mesons through flavor
singlet axial vector current. By using the pomeron-photon
analogy we can connect this matrix element with the matrix element
of the meson decay $M\rightarrow\gamma\gamma$. This decay is determined by
axial anomaly originated from a triangle graph
(see Fig.1). Therefore, by taking into account only difference
in the photon and pomeron coupling constants with quarks in the triangle
graph, we get the result
\be
p_M^\tau T_{\mu\nu\tau}=i\lambda F_{PPM}(t_1,t_2)\epsilon_{\mu\nu\rho\sigma}
q_1^\rho q_2^\sigma,
\label{anomaly}
\ee
where
\be
\lambda=\frac{18D\beta_0^2}{D^\prime\alpha}
\sqrt{\frac{2\Gamma_{\gamma\gamma}}{\pi M^3}}.
\la{coup}
\ee
Here  $\Gamma_{\eta\rightarrow\gamma\gamma}=0.46\times10^{-6} GeV$,
$\Gamma_{\eta^\prime\rightarrow\gamma\gamma}=4.28\times10^{-6} GeV$ and
factors $D$ and $D^\prime$ are related with the
wave functions of the $\eta$ and $\eta^\prime$
\be
\eta=-sin\Theta\eta_0+cos\Theta\eta_8 {\ }{\ }
\eta^\prime=cos\Theta\eta_0+sin\Theta\eta_8,
\la{wf}
\ee
\be
D_{\eta}=-sin\Theta,{\ }   {\ }   D_{\eta^\prime}=cos\Theta
\nn
\ee
\be
D_{\eta}^\prime=2\sqrt{2}cos\Theta-sin\Theta, {\ } {\ }
D_{\eta^\prime}^\prime=2\sqrt{2}cos\Theta+sin\Theta,
\nn
\ee
where $\Theta=-19.5^\circ$ is a singlet-octet mixing angle.
At high energies we have
\be
\bar u(p_{1,2}^\prime)\gamma_\mu u(p_{1,2})\approx
(p_{1,2}+p_{1,2}^\prime)_\mu,
\nn
\ee
and thus, the matrix element \re{matrix} becomes
\be
T=i36\beta_0^2\lambda(S_1/S_0)^{\alpha_P(t_1)-1}
(S_2/S_0)^{\alpha_P(t_2)-1}\epsilon_{\mu\nu\rho\sigma}p_1^\mu p_2^\nu
p_1^{\prime\rho}p_2^{\prime\sigma}.
\la{matfinal}
\ee
By using \re{ps2} and \re{matfinal} and the equations
\be
{\vec{p}_{1,2T}}^2=-x_{1,2}t_{1,2}-(1-x_1)^2m_p^2
\la{pt}
\ee
which follow from \re{kinematic},
the cross section is finally described by
\ba
\frac{d\sigma}{dt_1dt_2dx_Fd\Phi}&=&\frac{3^4\beta_0^4\lambda^2F_p^2(t_1)F_p^2(t_2)
F_{PPM}^2(t_1,t_2)}{2^9\pi^4\sqrt{x_F^2+4M^2/S}}(x_1t_1+(1-x_1)^2m_p^2)
(x_2t_2+(1-x_2)^2m_p^2)\nn\\
&\cdot& (S_1/S_0)^{2(\alpha_P(t_1)-1)}
(S_2/S_0)^{2(\alpha_P(t_2)-1)}sin^2\Phi,
\la{crossf}
\ea
where $x_1$ and $x_2$ are related with $x_F=x_2-x_1$ by the equation
$(1-x_1)(1-x_2)=M^2/S$ and
\be
S_{1,2}=S(1-x_{1,2})+m_p^2+2t_{1,2}.
\la{s12}
\ee
The kinematical limits for the phase space integration in \re{crossf} are
given from the positivity of ${\vec{p}_{1,2T}}^2$ \re{pt} and the condition
$S_{1,2}\geq (M+m_p)^2$.

It should be mentioned that the cross section \re{crossf} has a specific
azimuthal angle dependence which is related with the Lorenz structure of the
matrix element of axial anomaly \re{anomaly}. The cross section has a
maximum at $90^\circ$. The same dependence has been observed
by WA102 Collaboration \ct{WA102}.

The cross sections for the $\eta$ and $\eta^\prime$ production
were calculated for WA102 kinematics at $\sqrt{S}=29.1 GeV$.
The total cross sections calculated in the interval $0\leq x_F\leq 0.1$
are
\footnote{ We also have performed the exact phase space integration
for the DPE reaction without using high energy approximation \re{ps2}.
 The final result is slightly smaller than \re{res1}.}
\be
\sigma(\eta)=49 nb, {\ }{\ } {\ } {\ } {\ } \sigma(\eta^{\prime})=422 nb,
\la{res1}
\ee
which should be compared with the experiment data \ct{WA102}
\be
\sigma(\eta)^{exp}=1295\pm 16\pm 120 nb, {\ } {\ } \sigma(\eta^{\prime})^{exp}=
588\pm 18\pm 60 nb.
\la{exp}
\ee

Considering uncertainty of the pomeron-quark coupling constant
and experimental errors, the DPE model seems to be not inconsistent
with data of $\eta^\prime$ productions.  On the contrary, the situation
is hopeless for $\eta$ productions: DPE cannot explain at all
so big experimental data of cross sections.
It should be noticed that the small DPE contribution to $\eta$
meson productions is related to the structure of its wave function
\re{wf} which contains only a small flavor singlet component that
can contributes to DPE. It is impossible to obtain
$\sigma(\eta)>\sigma(\eta^\prime)$  for any mechanism of
these mesons productions which are sensitive only to the
flavor singlet component of their wave function.

As for $\eta^\prime$ productions, the DPE model seems to work rather well.
The $x_F$ dependence of the DPE cross section for $\eta^\prime$
productions is in qualitative agreement with experimental data,
though the cross section is decreasing faster than the data at large
$x_F$ regions, as shown in Fig.2.  The differential cross section
of $\eta^\prime$ productions is also presented in Fig.3 as a function
of the momentum transfer from one of the proton vertices.
This dependence is in  agreement with
experimental data which shows fast decreasing of the production at
large $t$.
\footnote{ The $t-$ dependence of $\eta$ and $\eta^\prime$
productions is very sensitive to the form factor in the
pomeron-pomeron-meson vertex. The statement that DPE should have
$e^{-b|t|}$ behavior \ct{WA102} is not correct, because one should
also take into account an additional $t-$dependence connected
with nonlocality of the pomeron-pomeron-meson vertex \re{PPM}.}.

\begin{figure}[htb]
\centering
\begin{minipage}[c]{7.5cm}
\vskip -.5cm

\centering
\epsfig{file=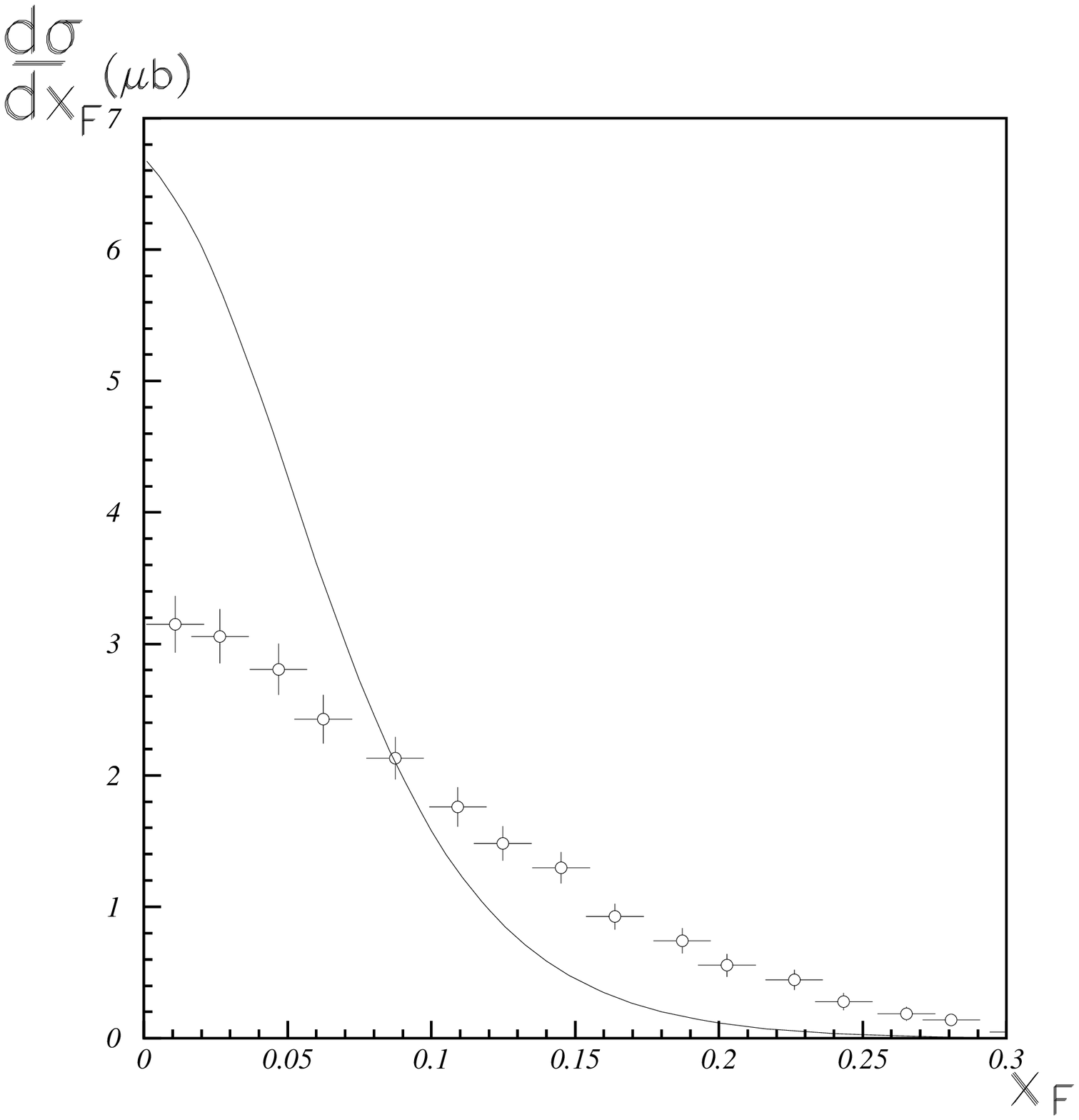,width=7.5cm}
\caption{\it $x_F$-dependence of the DPE contribution to
 $\eta^\prime$  central production in comparison with
experimental data of WA102 Collaboration. The experimental data
have been normalized to the total DPE cross section.}
\end{minipage}
%\vskip -1cm
\hspace*{11mm}
\begin{minipage}[c]{7.5cm}
\centering
\vskip -1.cm
\epsfig{file=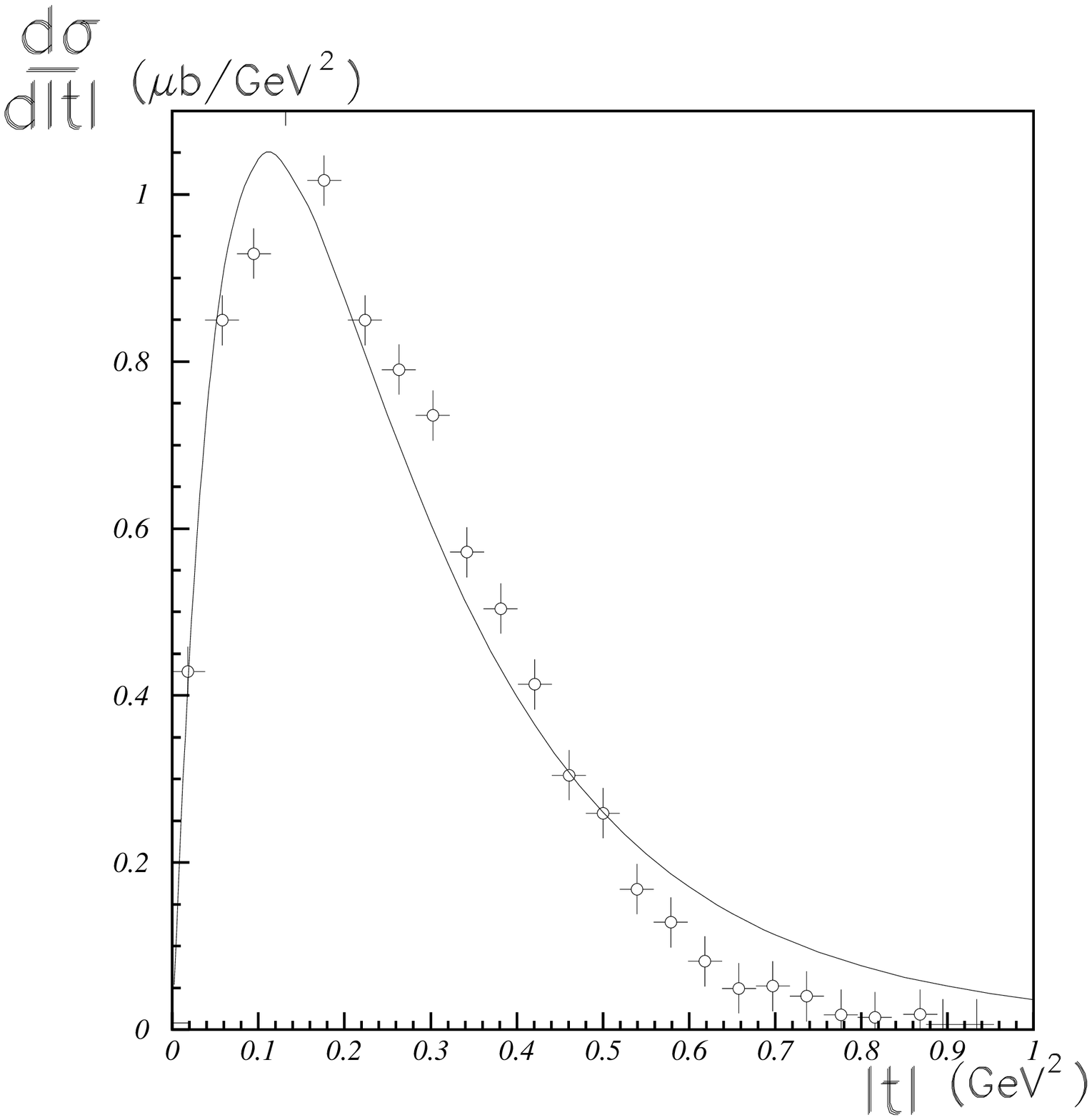,width=7.5cm}
\caption{\it $t$- dependence of the $\eta^\prime$
the  central production. The normalization of the experimental
data is the same as in Fig.2.
\protect\rule[-\baselineskip]
 {0pt}{2\baselineskip} }
\end{minipage}
\end{figure}

The energy dependence of the $\eta$ production
was measured  by WA102 Collaboration \ct{WA102E} and was found that it
 decreased with energy.
 This fact is also in contradiction with prediction
of slightly increasing of the DPE cross section with energies \re{crossf}.

In summary, although the DPE model
seems to work rather well for $\eta^\prime$ productions, it does not
explain the $\eta$ productions.  The situation is rather complicated:
the full set of the WA102 data at $\sqrt{S}=29.1 GeV$ cannot be explained
by the DPE model alone and we need some other mechanisms.  One of them
can be related with the contribution of the
nonperturbative fluctuation of gluon fields, i.e. instantons to
the central meson productions \ct{koch}.

 The authors are very grateful to  A.E.~Dorokhov, S.B.~Gerasimov,
and V.~Vento for useful discussions.

\end{document}